\documentclass[aps,prb,twocolumn,superscriptaddress]{revtex4-1}

\usepackage[dvipdf]{graphicx}

\usepackage{amssymb}

\usepackage{amsbsy}

\usepackage{latexsym}

\usepackage{multirow}

\begin{document}

\title{Thermal conductivity of suspended pristine graphene measured by Raman spectroscopy}

\author{Jae-Ung \surname{Lee}}
\affiliation{Department of Physics, Sogang University, Seoul 121-742, Korea}

\author{Duhee \surname{Yoon}}
\affiliation{Department of Physics, Sogang University, Seoul 121-742, Korea}

\author{Hakseong \surname{Kim}}
\affiliation{Division of Quantum Phases and Devices, School of Physics, Konkuk University, Seoul 143-701, Korea}

\author{Sang Wook \surname{Lee}}
\affiliation{Division of Quantum Phases and Devices, School of Physics, Konkuk University, Seoul 143-701, Korea}

\author{Hyeonsik \surname{Cheong}}
\email{hcheong@sogang.ac.kr}
\affiliation{Department of Physics, Sogang University, Seoul 121-742, Korea}

\date{\today}

\begin{abstract}
The thermal conductivity of suspended single-layer graphene was measured as a function of temperature using Raman scattering spectroscopy on clean samples prepared directly on a prepatterned substrate by mechanical exfoliation without chemical treatments. The temperature at the laser spot was monitored by the frequency of the Raman 2$D$ band of the Raman scattering spectrum, and the thermal conductivity was deduced by analyzing heat diffusion equations assuming that the substrate is a heat sink at ambient temperature. The obtained thermal conductivity values range from $\sim$1800~Wm$^{-1}$K$^{-1}$ near 325~K to $\sim$710~Wm$^{-1}$K$^{-1}$ at 500~K.
\begin{description}
\item[PACS numbers]
65.80.Ck, 63.22.Rc, 78.67.Wj.
\end{description}
\end{abstract}

\maketitle

\section{INTRODUCTION}

 Graphene is attracting much interest not only owing to its novel physical properties,\cite{GeimRev,KatsnelsonRev,NetoRev} but also because of possible applications as a candidate material to replace silicon in future electronic devices.\cite{GeimRev2,AvourisRev,FerraiRev} In addition to its high charge carrier mobility,\cite{Morozov,Bolotin} its superior thermal properties are considered to be crucial in high-density large-scale integrated circuits where heat management is becoming more important as the density of devices grows.\cite{Pop} Balandin \textit{et al.} first reported extremely large values for the thermal conductivity ($\kappa$) in the range of $4840\pm440$ to $5300\pm480$~Wm$^{-1}$K$^{-1}$ for mechanically exfoliated single-layer graphene near room temperature.\cite{Balandin} These values are among the largest ever measured from any material so far.

 Several groups since have measured $\kappa$ of mechanically exfoliated\cite{Balandin,Faugeras,Seol} or chemical vapor deposition (CVD)-grown\cite{Cai,Chen} graphene samples using different methods. For \emph{suspended}, exfoliated single-layer graphene, Faugeras \textit{et al.} reported a value of $\sim$630~Wm$^{-1}$K$^{-1}$ at 660 K,\cite{Faugeras} much lower than those of Balandin \textit{et al.}\cite{Balandin} On CVD-grown graphene, Ruoff's group\cite{Chen,Cai} reported $\kappa$ values ranging from $(2500 +1100/-1050)$~Wm$^{-1}$K$^{-1}$ to (2600$\pm900$ to 3100$\pm1000$)~Wm$^{-1}$K$^{-1}$ at 350~K. Since $\kappa$ is in principle a function of temperature and the measured values may be affected by the residual chemicals left on the samples as a result of the sample preparation processes, direct comparison of these later values with those of Balandin \textit{et al.} has been difficult. Given the importance of this key parameter for device applications, an accurate measurement and critical comparison with previous measurements are crucial. Here, we present the measurement of $\kappa$ for \emph{suspended} single-layer graphene at temperatures between 300~K and 500~K using Raman scattering spectroscopy on a clean sample prepared directly on a prepatterned substrate without involving a transfer process.

\section{EXPERIMENTAL}

\begin{figure}[b!]
\includegraphics{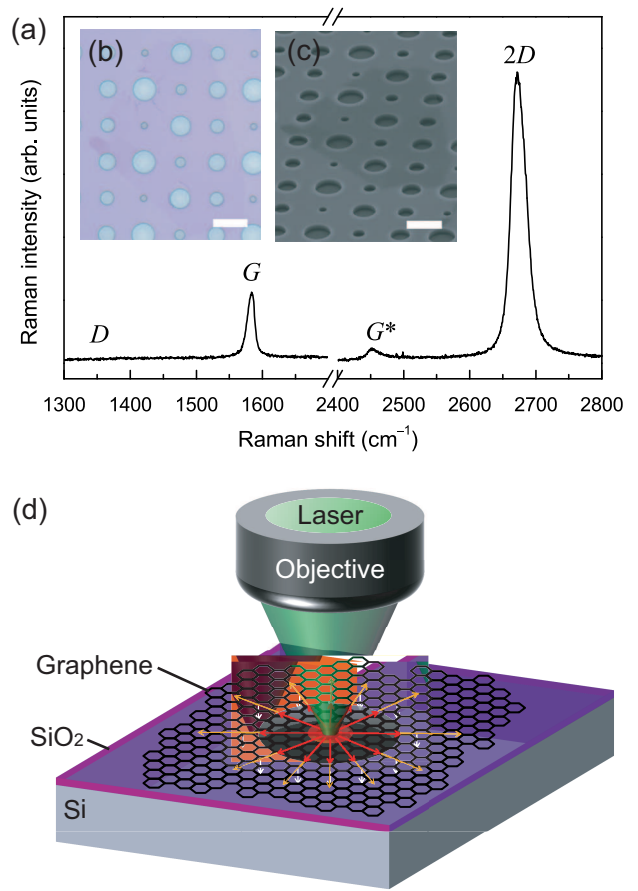}
\caption{(Color online) (a) Raman spectrum of suspended graphene. (b) Optical microscope and (c) scanning electron microscopy images of suspended graphene sample. The scale bars are 10 $\mu$m. (d) Schematic diagram of the experiment.}
\label{fig:fig1.eps}
\end{figure}

The substrates with round holes with various diameters were prepared by photolithography and dry etching of Si substrates covered with a 300~nm-thick SiO$_{2}$ layer. The depth of the holes is $\sim$1.7~$\mu$m, deep enough to prevent interference from laser light reflected and scattered from the bottom of the holes.\cite{Yoon} The diameters were 2.6, 3.6, 4.6, and 6.6~$\mu$m. The graphene samples were prepared directly on the cleaned substrate by mechanical exfoliation from natural graphite flakes. No chemical treatment of the sample was involved in the preparation process. This ensures that the sample surface is free from chemical contaminants that may affect the measured $\kappa$ values. The sample used was a single-layer graphene flake of 35$\times$60~$\mu$m$^{2}$ dimensions identified by the line shape of the $2D$ band in the Raman spectrum\cite{Ferrari,Yoon2} (Fig.\ref{fig:fig1.eps}). The 514.5-nm (2.41~eV) beam of an Ar ion laser was focused onto the graphene sample by a 50$\times$ microscope objective lens (0.8 N.A.), and the scattered light was collected and collimated by the same objective. The scattered signal was dispersed with a Jobin-Yvon Triax 550 spectrometer (1800 grooves/mm) and detected with a liquid-nitrogen-cooled charge-coupled-device detector. The spectral resolution was about 0.7~cm$^{-1}$. The laser spot size was measured using the modified knife-edge method:\cite{Cai,Veshapidze} the Raman intensity of the Si phonon peak was monitored as the laser spot is scanned across the straight sharp edge of a Ti patch deposited on Si. By fitting the intensity to $I(r)=I_0e^{-2r^2/w^2}$, $w = 0.29~\mu$m was obtained. Figure \ref{fig:fig1.eps} shows a typical Raman spectrum of suspended single-layer graphene obtained with a laser intensity of 1.0~mW. There is no indication of the defect-induced $D$ peak, attesting to the high quality of the sample. Although all our measurements were performed on a single piece of a graphene sample, there are some hole-to-hole variations in the low-power Raman spectrum, indicating some inhomogeneities.

\section{RESULTS AND DISCUSSION}

The absorption of the laser beam by the sample induces local heating that raises the temperature in the vicinity of the laser spot. In a steady state, there exists a temperature gradient that depends on the total power supplied by the laser beam, $\kappa$, and the boundary conditions at the edge of the hole. The local temperature at the laser spot can be estimated from the shift of the Raman $G$ or $2D$ bands. The temperature dependence of the Raman spectrum of graphene has been studied by several groups.\cite{Cai,Chen,Calizo,Allen,Jo} Most of the studies were conducted on graphene samples on substrates. In those cases, the Raman spectrum may be affected by the strain induced by the difference in the thermal expansion coefficients of the substrate and graphene, in addition to the purely thermal effect. Since graphene samples suspended over a trench or a hole is less affected by the strain due to the thermal expansion coefficient difference, we used the values reported recently by Chen \textit{et al.}\cite{Chen} on \emph{suspended} graphene samples. They measured the temperature coefficients of the $G$ and $2D$ bands and found that the $2D$ band ($\partial\omega_{2D}/{\partial}T=-0.072\pm0.002$cm$^{-1}$K$^{-1}$) is more sensitive to temperature than the $G$ band ($\partial\omega_{G}/{\partial}T=-0.044\pm0.003$cm$^{-1}$K$^{-1}$). Therefore, we used the $2D$ band for the estimate of the temperature in this work.

\begin{figure}
\includegraphics{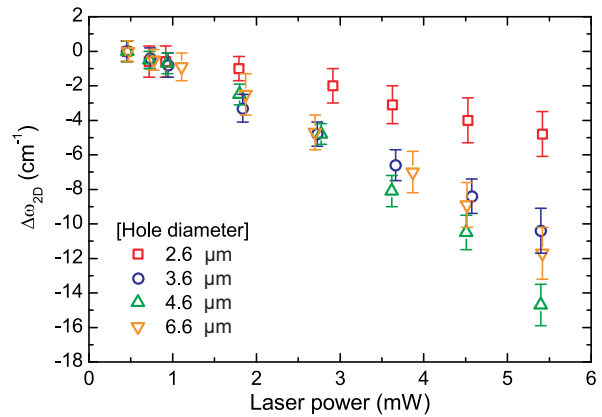}
\caption{(Color online) Shift of the Raman $2D$ band as a function of the laser power.}
\label{fig:fig2.eps}
\end{figure}

Figure \ref{fig:fig2.eps} summarizes the shift of the $2D$ band as a function of the incident laser power. As the laser power increases, the $2D$ band frequency redshifts due to increased heating. For smaller holes, the shift is smaller because efficient heat conduction to the substrate limits the temperature rise at the laser spot. It should be noted that the largest hole (6.6~$\mu$m) shows a smaller shift than the 4.6-$\mu$m hole. This trend was confirmed by repeated measurements on several holes. This may be explained in the following way. As the hole size increases, the laser spot moves away from the edge of the hole, reducing the heat conduction to the substrate, and the temperature should increase. Beyond a certain hole size, the temperature would saturate if one ignores the thermal conduction to the ambient air. In reality, the conduction to air, however small, would decrease the temperature for larger holes, resulting in a smaller temperature rise for the larger holes.

\begin{table*}
\caption{\label{tab:table1} Comparison on thermal conductivity of \emph{suspended} single-layer graphene.}
\begin{ruledtabular}
\begin{tabular}{c c c c c c}
 Sample type & Temp. determination & Shape & $\kappa$ (Wm$^{-1}$K$^{-1}$) & $T$ & Ref.\\ \hline
 Exfoliated, pristine & $2D$ band & Circular & $\sim$1800 & $\sim$325 K & This work \\
 &  &  & $\sim$710 & $\sim$500 K & \\
 Exfoliated, pristine & $G$ band & Trench & $\sim$4840--5300 & RT & Balandin \textit{et al.}\cite{Balandin} \\
 Exfoliated, transfered & Stokes/anti-Stokes & Circular & $\sim$630 & $\sim$660 K & Faugeras \textit{et al.}\cite{Faugeras} \\
 CVD, transfered & $G$ band & Circular & $\sim$2500 & $\sim$350 K & Cai \textit{et al.}\cite{Cai} \\
 &  &  & $\sim$1400 & $\sim$500 K &  \\
 CVD, transfered & $2D$ band & Circular & $\sim$2600--3100 & $\sim$350 K & Chen \textit{et al.}\cite{Chen} \\
\end{tabular}
\end{ruledtabular}
\end{table*}

In order to estimate the thermal conductivity, we used the heat diffusion equation ignoring the heat conduction to the ambient air. We considered heat conduction through suspended graphene and supported graphene on the substrate as well as between graphene and the substrate. The substrate is assumed to be a heat sink at the ambient temperature. With cylindrical symmetry, one can write the heat diffusion equation as:

\begin{equation}\label{eq:1}
\kappa\frac{1}{r}\frac{d}{dr}\Big(r\frac{dT_1(r)}{dr}\Big)+q(r)=0\quad \textrm{for}\ r < R,
\end{equation}

\noindent
where $R$ is the radius of the hole, $r$ is the radial position, and $\kappa$ is the thermal conductivity of \emph{suspended} graphene. $q(r)=(I\alpha/t)exp(-2r^2/w^2)$ is the heat inflow per unit volume due to laser excitation, where $I$ is the laser intensity, $\alpha$ is the absorptance of light in single layer graphene (2.3\%),\cite{Nair,Stauber,Mak,Sheehy} and $t$ is the thickness of graphene (0.335~nm). Outside the hole, where graphene is supported by the substrate, the following equation applies:

\begin{equation}\label{eq:2}
\kappa^\prime\frac{1}{r}\frac{d}{dr}\Big(r\frac{dT_2(r)}{dr}\Big)-\frac{\sigma_i}{t}(T_2(r)-T_a)=0\quad \textrm{for}\ r \geq R,
\end{equation}

\noindent
where $\kappa^\prime$ is the thermal conductivity of \emph{supported} graphene (600~Wm$^{-1}$K$^{-1}$),\cite{Seol} $T_a$ is the ambient temperature, and $\sigma_i$ is the interfacial thermal conductance between graphene and SiO$_2$ (100~MWm$^{-2}$K$^{-1}$).\cite{Chen2} The general solutions to Eqs.~(\ref{eq:1}) and (\ref{eq:2}) are

\begin{eqnarray}
\label{eq:3}&&T_1(r)=c_1+c_2 \ln(r)+c_3Ei\Big(\frac{-r^2}{r_0^2}\Big)\quad \textrm{for}\ r < R, \\
\label{eq:4}&&T_2(\gamma)=c_4I_0(\gamma)+c_5K_0(\gamma)+T_a\quad \textrm{for}\ r \geq R,
\end{eqnarray}

\noindent
where $c_i$'s are arbitrary constants, $Ei(x)$ is an exponential integral, $I_0(x)$ and $K_0(x)$ are the zero-order modified Bessel functions, and $\gamma=r(\sigma_i/(\kappa^\prime t))^\frac{1}{2}$. For a converging solution, $c_4$=0. The boundary conditions are:

\begin{eqnarray}
& &T_2(r\to\infty)=T_a,  \\
& &T_1(R)=T_2(\gamma)\vert_{r=R}, \\
& &-\kappa\frac{dT_1(r)}{dr}\vert_{r=R}=-\kappa^\prime \frac{dT_2(\gamma)}{dr}\vert_{r=R},   \\
& &-2{\pi}Rt\kappa^\prime\frac{dT_2(\gamma)}{dr}\vert_{r=R}=Q,
\end{eqnarray}

\noindent
where $Q$ is the total laser power absorbed. The coefficients $c_i$'s of Eqs.~(\ref{eq:3}) and (\ref{eq:4}) are determined from these boundary conditions. On the other hand, the measured temperature ($T_m$) is an weighted average of temperature inside the beam spot and can be approximated as

\begin{equation}\label{eq:9}
T_m\approx\frac{\int_{0}^{w}T_1(r)q(r)rdr}{\int_{0}^{w}q(r)rdr}.
\end{equation}

\begin{figure} [t]
\includegraphics{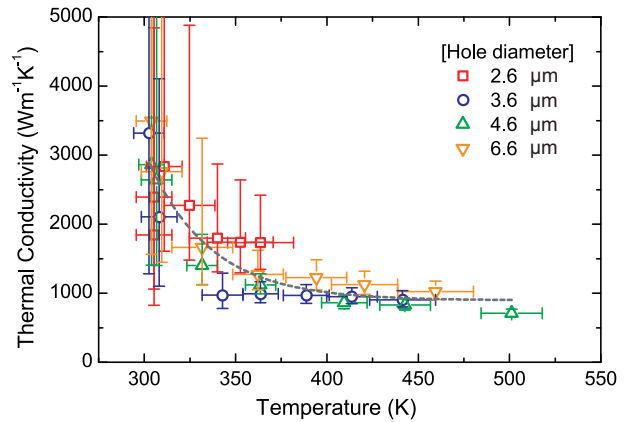}
\caption{(Color online) Thermal conductivity of suspended graphene as a function of measured temperature. The dotted curve is guide to eye.}
\label{fig:fig3.eps}
\end{figure}

By comparing the measured $T_m$ with Eq.~(\ref{eq:9}), one can determine the thermal conductivity $\kappa$. Figure \ref{fig:fig3.eps} shows thus determined $\kappa$ as a function of the measured temperature. The error bars are quite large for lower temperatures because $\Delta\omega_{2D}$, which determines $T_m$, is quite small in comparison to the measurement resolution. Therefore, the $\kappa$ values at temperatures below 325 K are not very reliable. It seems that $\kappa$ decreases as the temperature increases: from $\sim$1800~Wm$^{-1}$K$^{-1}$ near 325~K to $\sim$710~Wm$^{-1}$K$^{-1}$ near 500~K.

Our thermal conductivity values are somewhat lower than those reported for CVD graphene.\cite{Chen} In that work, the measured $\kappa$ is 2600 -- 3100~Wm$^{-1}$K$^{-1}$ near 350~K. The major difference in that work is that they measured the transmittance ($I_t/I_0$) of the graphene sample and took $1-I_t/I_0$ as the absorptance. Their value was 3.4\%, which is 50\% larger than the recently determined value of 2.3\%.\cite{Nair,Stauber,Mak,Sheehy} Reflection and scattering of light on the sample surface and/or other loss of the transmitted light may account for the difference. If we use the absorptance of 3.4\%, we obtain a $\kappa$ value of $\sim$2700~Wm$^{-1}$K$^{-1}$ at $\sim$325~K, similar to the value for CVD-grown graphene. On the other hand, Faugeras \textit{et al.} reported $\kappa$ of $\sim$630~Wm$^{-1}$K$^{-1}$ at $\sim$660~K for exfoliated graphene,\cite{Faugeras} which is rather close to our result at 500~K. Another source of uncertainty in the analysis is the temperature coefficient of the Raman $2D$ band. A 20\% variation in the temperature coefficient value would result in a similar variation in the obtained $\kappa$ value.


In light of the above analysis, the initially reported value of 5300~Wm$^{-1}$K$^{-1}$ by Balandin \textit{et al.} seems to be significantly overestimated. The most significant difference between their analysis and those of recent publications including our work is the value of the absorptance $\alpha$ of single layer graphene. They used $\alpha=13$\%, which is several times larger than the  value of 2.3\% accurately measured and theoretically analyzed by Nair \textit{et al}.\cite{Nair} If one uses $\alpha=2.3$\%, their $\kappa$ value would reduce to $\sim940$~Wm$^{-1}$K$^{-1}$.

\section{CONCLUSIONS}
The thermal conductivity of suspended single-layer graphene was measured as a function of temperature using Raman spectroscopy on pristine graphene samples prepared directly on a patterned substrate by mechanical exfoliation. By monitoring the temperature at the laser spot using the Raman $2D$ band, the thermal conductivity was deduced by analyzing heat diffusion equations. The obtained thermal conductivity values range from $\sim$1800~Wm$^{-1}$K$^{-1}$ near 325~K to $\sim$710~Wm$^{-1}$K$^{-1}$ near 500~K. Based on our result as well as other recent reports,\cite{Faugeras,Chen} the initially reported\cite{Balandin} value of 5300~Wm$^{-1}$K$^{-1}$ seems to be significantly overestimated.

\begin{acknowledgments}
This work was supported by Mid-career Researcher Program through NRF grant funded by the MEST (No. 2008-0059038). D.~Y.~acknowledges funding from the Seoul City Government. H.~K.~and S.~L.~are supported by WCU program through the NRF funded by the MEST (No. R31-2008-000-10057-0)
\end{acknowledgments}


\begin{thebibliography}{99}

\bibitem{GeimRev}A. K. Geim and K. S. Novoselov, Nat. Mater. \textbf{6}, 183 (2007).

\bibitem{KatsnelsonRev}M. I. Katsnelson, Mater. Today \textbf{10}, 20 (2007).

\bibitem{NetoRev} A. H. Castro Neto, F. Guinea, N. M. R. Peres, K. S. Novoselov, and A. K. Geim, Rev. Mod. Phys. \textbf{81}, 109 (2009).

\bibitem{GeimRev2} A. K. Geim, Science \textbf{324}, 1530 (2009).

\bibitem{AvourisRev} Ph. Avouris, Nano Lett. \textbf{10}, 4285 (2010).

\bibitem{FerraiRev} F. Bonaccorso, Z. Sun, T. Hasan, and A. C. Ferrari, Nat. Photonics \textbf{4}, 611 (2010).

\bibitem{Morozov} S. V. Morozov, K. S. Novoselov, M. I. Katsnelson, F. Schedin, D. C. Elias, J. A. Jaszczak, and A. K. Geim, Phys. Rev. Lett. \textbf{100}, 016602 (2008).

\bibitem{Bolotin} K. I. Bolotin, K. J. Sikes, J. Hone, H. L. Stormer, and P. Kim, Phys. Rev. Lett. \textbf{101}, 096802 (2008).

\bibitem{Pop} E. Pop, Nano Res. \textbf{3}, 147 (2010).

\bibitem{Balandin}A. A. Balandin, S. Ghosh, W. Bao, I. Calizo, D. Teweldebrhan, F. Miao, and C. N. Lau, Nano Lett. \textbf{8}, 902 (2008); S. Ghosh, I. Calizo, D. Teweldebrhan, E. P. Pokatilov, D. L. Nika, A. A. Balandin, W. Bao, F. Miao, and C. N. Lau, Appl. Phys. Lett. \textbf{92}, 151911 (2008).

\bibitem{Faugeras}C. Faugeras, B. Faugeras, M. Orlita, M. Potemski, R. R. Nair, and A. K. Geim, ACS Nano \textbf{4}, 1889 (2010).

\bibitem{Seol}J. H. Seol, I. Jo, A. L. Moore, L. Lindsay, Z. H. Aitken, M. T. Pettes, X. Li, Z. Yao, R. Huang, D. Broido, N. Mingo, R. S. Ruoff, and L. Shi, Science \textbf{328}, 213 (2010).

\bibitem{Cai}W. Cai, A. L. Moore, Y. Zhu, X. Li, S. Chen, L. Shi, and R. S. Ruoff, Nano Lett. \textbf{10}, 1645 (2010).

\bibitem{Chen}S. Chen, A. L. Moore, W. Cai, J. W. Suk, J. An, C. Mishra, C. Amos, C. W. Magnuson, J. Kang, L. Shi, and R. S. Ruoff, ACS Nano \textbf{5}, 321 (2011).

\bibitem{Yoon}D. Yoon, H. Moon, Y. W. Son, J. S. Choi, B. H. Park, Y. H. Cha, Y. D. Kim, and H. Cheong, Phys. Rev. B \textbf{80}, 125422 (2009).

\bibitem{Ferrari}A. C. Ferrari, J. C. Meyer, V. Scardaci, C. Casiraghi, M. Lazzeri, F. Mauri, S. Piscanec, D. Jiang, K. S. Novoselov, S. Roth, and A. K. Geim, Phys. Rev. Lett. \textbf{97}, 187401 (2006).

\bibitem{Yoon2}D. Yoon, H. Moon, H. Cheong, J. S. Choi, J. A. Choi, and B. H. Park, J. Korean Phys. Soc. \textbf{55}, 1299 (2009).

\bibitem{Veshapidze}G. Veshapidze, M. L. Trachy, M. H. Shah, and B. D. DePaola, Appl. Opt. \textbf{45}, 8197 (2006).

\bibitem{Calizo}I. Calizo, A. A. Balandin, W. Bao, F. Miao, and C. N. Lau, Nano Lett. \textbf{7}, 2645, (2007); I. Calizo, F. Miao,  W. Bao, C. N. Lau, A. A. Balandin, Appl. Phys. Lett. \textbf{91}, 071913, (2007).

\bibitem{Allen}M. J. Allen, J. D. Fowler, V. C. Tung, Y. Yang, B. H. Weiller, and R. B. Kaner, Appl. Phys. Lett. \textbf{93}, 193119 (2008).

\bibitem{Jo}I. Jo, I. Hsu, Y. J. Lee, M. M. Sadeghi, S. Kim, S. Cronin, E. Tutuc, S. K. Banerjee, Z. Yao, and L. Shi, Nano Lett. \textbf{11}, 85 (2011).

\bibitem{Nair}R. R. Nair, P. Blake, A. N. Grigorenko, K. S. Novoselov, T. J. Booth, T. Stauber, N. M. R. Peres, and A. K. Geim, Science \textbf{320}, 1308 (2008).

\bibitem{Stauber}T. Stauber, N. M. R. Peres, and A. K. Geim, Phys. Rev. B \textbf{78}, 085432 (2008).

\bibitem{Mak}K. F. Mak, M. Y. Sfeir, Y. Wu, C. H. Lui, J. A. Misewich, and T. F. Heinz, Phys. Rev. Lett. \textbf{101}, 196405 (2008).

\bibitem{Sheehy}D. E. Sheehy and J. Schmalian, Phys. Rev. B \textbf{80}, 193411 (2009).

\bibitem{Chen2}Z. Chen, W. Jang, W. Bao, C. N. Lau, and C. Dames, Appl. Phys. Lett. \textbf{95}, 161910 (2009).

\end{thebibliography}
\end{document}